\documentclass[a4paper,usenatbib]{mnras}

\usepackage{graphicx}
\usepackage{amsmath}
\usepackage{siunitx}
\usepackage{xcolor}

\usepackage{bm}

\usepackage[T1]{fontenc}
\usepackage{ae,aecompl}
\usepackage{times}

\usepackage{multirow}

\usepackage{newtxtext,newtxmath}

\usepackage{xspace}

\newcommand{\unitflux}{\,erg\,cm$^{-2}$\,s$^{-1}$}
\newcommand{\unitlumi}{\,erg\,s$^{-1}$}
\newcommand{\mission}[1]{\textit{#1}}
\newcommand{\msun}{$\mathrm{M}_\odot$\xspace}

\newcommand{\nhost}{$N_\mathrm{H, host}$\xspace}
\newcommand{\ngc}{NGC 7589\xspace}
\newcommand{\eddratio}{$\lambda_{\mathrm{Edd}}$\xspace}

\newcommand{\fluxfull}{f_\mathrm{0.5-10\,keV}}

\title[Large Amplitude Variability in NGC 7589]{The Large Amplitude X-ray Variability in NGC 7589: Possible Evidence for Accretion Mode Transition}

\author[Z. Liu]{Zhu Liu,$^{1}$\thanks{Contact e-mail: \href{mailto:liuzhu@nao.cas.cn}{liuzhu@nao.cas.cn}}
    He-Yang Liu,$^{1,2}$
    Huaqing Cheng,$^{1,2}$
    Erlin Qiao,$^{1,2}$
    Weimin Yuan$^{1,2}$
\\
$^{1}$ Key Laboratory of Space Astronomy and Technology, National Astronomical Observatories, Chinese Academy of Sciences, Beijing 100101, China \\
$^{2}$ University of Chinese Academy of Sciences, School of Astronomy and Space Science, Beijing 100049, China}

\date{}
\pubyear{2018}

\begin{document}
    
\label{firstpage}
\pagerange{\pageref{firstpage}–\pageref{lastpage}}
\maketitle


\begin{abstract}
We report the discovery of large amplitude X-ray variability in the low luminosity AGN (LLAGN) \ngc, and present possible observational evidence for accretion mode transition in this source. Long-term X-ray flux variations by a factor of more than 50 are found using X-ray data obtained by \mission{Swift/XRT} and \mission{XMM--Newton} over 17 years. Results of long-term monitoring data in the UV, optical and infrared bands over $\sim20$ years are also presented. The Eddington ratio \eddratio increased from $10^{-3}$ to $\sim0.13$, suggesting a transition of the accretion flow from an ADAF to a standard thin accretion disc. Further evidence supporting the thin disc in the high luminosity state is found by the detection of a significant soft X-ray component in the X-ray spectrum. The temperature of this component ($\sim19^{+15}_{-7}$\,eV, fitted with a blackbody model) is in agreement with the predicted temperature of the inner region for a thin disc around a black hole (BH) with mass of $\sim10^{7}$\msun. These results may indicate that NGC 7589 had experienced accretion mode transition over a timescale of a few years, suggesting the idea that similar accretion processes are at work for massive black hole and black hole X-ray binaries.
\end{abstract}


\section{Introduction}

It is believed that Active Galactic Nuclei (AGNs) are powered by accreting material onto massive black holes (BHs) at the centre of galaxies. Observational evidence of an optical thick, geometrical thin standard accretion disc \citep[hereafter thin disc,][]{shakura_sunyaev1973} has been found in the multi-band data of luminous AGNs (e.g. Seyferts and QSOs). For instance, the big blue bump observed in the spectral energy distribution (SEDs) of AGNs is explained as the thermal radiation from the thin disc \citep[e.g.][]{elvis_etal1986, cerny_elvis1987, shang_etal2005}. In addition, the broad Fe K$\alpha$ line, which is generally thought to originate from the inner region of an accretion disc and broadened due to relativistic effects of the central BH \citep{fabian_etal1989, fabian_etal2000}, has been detected in the X-ray spectra of some AGNs (e.g. \citealt{tanaka_etal1995, deLaCallePerez_etal2010}), supporting the thin disc model with an inner radius extended to a few gravitational radius ($R_{\mathrm{g}}=GM/c^2$) to the BH. The high energy tail of the thermal radiation from the thin disc, which can be modelled by a blackbody with temperature of several tens of eV (depends on the BH mass), has also been discovered in the X-ray spectra of some AGNs\footnote{However, the spectra in the soft X-ray band of the vast majority of AGN show a component with a higher temperature ($\sim0.1-0.2$\,keV) than the prediction of the thin disc -- the so-called soft X-ray excess whose origin is still under debate \citep[e.g.][]{arnaud1996, gierlinski_done2004}} \citep{yuan_etal2010, sun_etal2013, shu_etal2017}. On the other hand, an optical thin radiation inefficient accretion flow (RIAF) or advection dominated accretion flow (ADAF, \citealt{narayan_yi1994, abramowicz_etal1995, yuan_narayan2014}) has successfully explained the observed property of low luminosity AGNs (e.g. LLAGNs and LINERs, \citealt{ho2008}) that normally has very low accretion rates.

Theoretical calculation suggests that transition between different accretion mode will take place once the accretion rate reaches certain critical values \citep{meyer_etal2000}. Evidence for accretion mode transition has long been found in Galactic X-ray binaries (XRBs), e.g. \citet{zhang_etal1997}. In the high/soft state, a thermal component with temperature of $\sim1$\,keV seen in the soft X-ray band is widely accepted as the thermal radiation from the thin disc. While in the low/hard state, the hard X-ray spectrum are consistent with the prediction from ADAF \citep[e.g.][]{esin_etal1997}. Tentative evidence for accretion mode transition has been reported in a few AGNs \citep[e.g.][]{quataert_etal1999, yuan_narayan2004, xu_cao2009, xie_etal2016}. However, strong observational evidence for accretion mode transition in individual AGN has not yet been found, though it has been invoked to explain the large amplitude X-ray variability in a few AGNs (e.g. \ngc in \citealt{yuan_etal2004}, NGC\,7213 in \citealt{xie_etal2016}) and changing-look AGN \citep{noda_done2018}.

Observational evidence for accretion mode transition in AGNs is important not only because it proves that the accretion theory can be universally applied to stellar and supermassive BH--accretion system, but also because it may yield important insights into many yet unclear problems in AGNs. For instance, the expected SED from various accretion models are dramatically different, thus accretion mode transition can potentially explain the changing-look phenomenon found in some AGNs \citep[e.g.][]{macleod_etal2016, ruan_etal2016, yang_etal2018}. \citet{noda_done2018} found that the SED of the changing-look AGN Mrk 1018 is consistent with typical Seyfert 1 AGNs in the high state, while it can be modelled with ADAF in the low state. They thus proposed that accretion state transition can explain the changing-look phenomenon in AGNs. Moreover, such transition system can provide crucial clues, such as the critical accretion rate and the transition time-scale, to improve our understanding of the physical process that causes the transition.

\ngc, at a redshift of 0.0298, is optically classified as Seyfert 1.9/LINERs and a LLAGN based on the weak broad H{$\alpha$} line, and the BH mass is estimated to be $\sim10^7$\msun \citep{yuan_etal2004}. Using the archival \mission{Einstein}, \mission{ROSAT} and \mission{XMM--Newton} data, \citet{yuan_etal2004} found that this source showed large amplitude X-ray variability during the period from 1980 to 2001, i.e. the 0.5-2.4\,keV flux changed by a factor of 10 on a time-sale of months to years. The lowest Eddington ratio\footnote{{\eddratio}$ = L_{\rm bol}/L_{\rm Edd}$,
where $L_{\rm Edd} = 1.3\times10^{38}(M_{\rm BH}/M_{\odot})$ and $L_{\rm bol}$ is
the bolometric luminosity.} (the 1995 \mission{ROSAT} pointing observation) for this source was conservatively estimated to be less than $10^{-4}$ \citep{yuan_etal2004}, suggesting that the source was accreting via the RIAF/ADAF mode at its low luminosity state. While the accretion mode was not clear in the relatively high luminosity state observed in 2001 by \mission{XMM--Newton}, the estimated accretion rate was about a few percent, suggesting that the accretion may possibly be in a transition state between RIAF and the thin disc. \citet{yuan_etal2004} thus proposed that a transition of accretion state may have taken place in \ngc if the X-ray flux had ever reached a peak much higher than the flux state observed in 2001. However, as noted by the authors, a partial covering model which may also explain the observed variability cannot be ruled out.

In this work, using all the new serendipitous X-ray observations of \ngc from \mission{XMM--Newton} and the \mission{X-ray telescope} \citep{burrows_etal2005} onboard the \mission{Neil Gehrels Swift Observatory} (\mission{Swift/XRT}), we show that indeed \ngc can reach an even higher flux than the 2001 `high' luminosity state. A low luminosity state comparable to the 1995 \mission{ROSAT} observation was also detected in the 2018 \mission{Swift/XRT} observations. Results from the analysis of the X-ray spectra and multi-band variability suggest that the large amplitude variability is due to the change of intrinsic X-ray flux, rather than (partial covering) obscuration, and may indicate that a transition of accretion mode had taken place in \ngc. Throughout this paper, we adopted a flat $\Lambda\mathrm{CDM}$ cosmological model with $H_0=69.3\,\mathrm{km\,s^{-1}}$, $\Omega_m=0.29$ and $\Omega_\Lambda=0.71$. All the quoted uncertainties correspond to the 90 per cent confidence level for one interesting parameter, unless specified otherwise.

\section{Multi-band Data Reduction}
\ngc had been observed in X-ray band by \mission{Einstein} in 1985, \mission{ROSAT} in 1992 and 1995, \mission{XMM--Newton} in 2001 June and November, and 2006 June. It was also serendipitously observed by \mission{Swift/XRT} extensively from 2006 April till 2018 July. In this work, only the X-ray data from \mission{XMM--Newton} and \mission{Swift/XRT} observations were analysed (please refer to \citealt{yuan_etal2004} for the data analysis of X-ray observations taken before 2000). \ngc is located in the sky region covered by the SDSS Stripe 82 Survey. Optical and UV observations were also frequently carried out by \mission{SDSS} and \mission{GALEX} in the past decades, respectively. In this section, we described the X-ray, optical, UV, and MIR multi-band data analysis.

\begin{table*}
\caption{\label{tab:obs_log}\mission{XMM--Newton} and \mission{Swift} observation logs}
\begin{tabular}{lccccc}\hline
Obs ID      & Obs date                              & Exposure (s)                         & Instrument & Exposure (s)             & Optical/UV filters           \\\hline
            &                                       & 		   &   \mission{XMM--Newton} observations          &                      &                      \\
0066950301  & 2001-06-03                            & 7176/6961                            & M1/M2      & 2400                 & UVW1                 \\
0066950401  & 2001-11-28                            & 2897/7506/7999                       & pn/M1/M2   & 8000                 & UVW1                 \\
0305600601  & 2006-06-14                            & 15970/16020                          & M1/M2      & ---                  & ---                  \\
            &                                       & 	   	   &   \mission{Swift} observations        &                      &                      \\
00035365001 & 2006-01-15                            & 1730                                 & XRT        & ---                  & ---                  \\
00035365002 & 2006-04-29                            & 401                                  & XRT        & ---                  & ---                  \\
00049538001 & 2013-01-22                            & 928                                  & XRT        & ---                  & ---                  \\
00049538002 & 2013-04-28                            & 2031                                 & XRT        & ---                  & ---                  \\
00049538003 & 2013-09-23                            & 888                                  & XRT        & 72/72/72/144/218/288 & V/B/U/UVW1/UVM2/UVW2 \\
00049538004 & 2014-09-01                            & 341                                  & XRT        & 340                  & UVW2                 \\
00049538005 & 2014-10-29                            & 115                                  & XRT        & 340                  & UVW1                 \\
00049538006 & 2015-01-21                            & 702                                  & XRT        & ---                  &                      ---\\
00049538007 & 2015-09-21                            & 356                                  & XRT        & 349                  & UVM2                 \\
00049538008 & 2016-01-19                            & 1596                                 & XRT        & ---                  &                      ---\\
00049538009 & 2016-01-20                            & 3510                                 & XRT        & 829                  & UVW1                 \\
---         & 2018 combined                         & 25000                                & XRT        & 1138/1016/946/526    & UVW2/UVM2/UVM2/UVM2  \\ \hline
\end{tabular}
\end{table*}

\subsection{\mission{XMM--Newton}}
The Observation Data Files (ODF) were downloaded from the \mission{XMM--Newton} Science Archive. The ODFs were then reduced using the \mission{XMM-Newton} Science Analysis System (\textsc{SAS}) software \citep[version 16.1,][]{gabriel_etal2004}. The \textsc{SAS} tasks \textsc{emchain} and \textsc{epchain} were used to generate the event lists for the European Photon Imaging Camera (EPIC) MOS \citep{turner_etal2001} and pn \citep{struder_etal2001} detectors, respectively. High background flaring periods were then identified and filtered from the event lists. A circular region with radius of $32$ and $35$\,arcsec was selected as the source region for the two 2001 observations and the 2006 observations (the source was much brighter and off-axis), respectively. For the background region, an annulus (concentered with the source) with inner radius of $40$\,arcsec and outer radius of $120$\,arcsec were chosen for the 2001 EPIC MOS observations, while an circular region with radius of $70$\,arcsec was selected for the EPIC pn data (only available for the second 2001 \mission{XMM--Newton} observation). We selected an annular region with inner radius of $60$\,arcsec and outer radius of $140$\,arcsec for the 2006 EPIC MOS1 observation. In the case of MOS2, an annulus is impossible, thus a circle with radius of $100$\,arcsec was chosen as the background region. The \textsc{arfgen} and \textsc{rmfgen} tasks were used to generate the response files.

The OM data with the filter UVW1 were available for two observations (obsID:0066950301, 0066950401). There are four exposures (in total 2400\,s) in the first observation and 10 exposures in the second observations (total exposure time of 8000\,s). The \textsc{SAS} task \textsc{omichain} was used to generate the OM images. The photometric measurements were performed for each exposure in the two observations with the task \textsc{omphotom}. A circular region with radius of 6\,arcsec was selected for the source aperture, while an annulus with inner radius of 9 and outer radius of 14 was chosen to estimate the background. No significant variability was found within each observation, thus the mean magnitude of all the exposures in each observation was then used. The standard deviation was considered as the 1\,$\sigma$ uncertainty. The details of all the \mission{XMM--Newton} observations can be found in Table\,\ref{tab:obs_log}.

\subsection{\mission{Swift} observations}
\ngc was serendipitously observed 56 times by the \mission{Swift/XRT} from 2006 to 2018. The light curve as well as the X-ray spectra were generated using the \mission{XRT} online data analysis tools\footnote{\url{http://www.swift.ac.uk/user_objects}} \citep{evans_etal09}. The $3\,\sigma$ flux upper limits (assuming an absorbed power-law model, see Section\,\ref{sec:xrayhig_spec}) were given for observations that the source was not detected. \ngc was in a low X-ray luminosity state in 2018, resulted in low data quality of the X-ray spectra. To increase the S/N, we generated one combined X-ray spectrum using all the 2018 \mission{Swift/XRT} observations. The details of the \mission{Swift/XRT} observations can be found in Table\,\ref{tab:obs_log} (note that all the observations taken in 2018 were combined and named as 2018 combined data).

The \mission{Swift/UVOT} data were available for several observations (obsIDs: 00049538003, 00049538004, 00049538005, 00049538007, 00049538009, 00049538011, 00049538013, 00049538014, 00049538016). The source counts were extracted from a circular region with radius of 5\,arcsec, while a 15\,arcsec circle was choose as the background region. The task \textsc{uvotsource} was used to perform photometric measurements. The filters and exposure times in each observation can be found in Table\,\ref{tab:obs_log}.

\subsection{\mission{SDSS}}
\ngc is located in the sky area covered by the SDSS Stripe 82 Survey, and was frequently observed in $u, g, r, i, z$ bands with SDSS. We searched in the SDSS Stripe 82 Sky Survey at the position of \ngc with a searching radius of 0.5\,arcsec. This resulted in a total of 77 observations. The measured photometric magnitudes and observation time of the $u,g,r,i,z$ bands for each individual observation were obtained from the SDSS Stripe 82 Sky Survey Database\footnote{\url{http://cas.sdss.org/stripe82/en}}. We noted that the contribution from host galaxy was not subtracted.

\subsection{\mission{GALEX}}
\ngc was observed by \mission{Galaxy Evolution Explorer (GALEX)} 11 times with a total exposure time of 8874\,seconds in the NUV band, and 8 times with 6886\,second exposure time in the FUV band. The \textsc{gPhoton} \citep{million_etal2016a, million_etal2016b} Python package\footnote{\url{https://archive.stsci.edu/prepds/gphoton}} was adopted to analyze the \mission{GALEX} NUV and FUV data. Intensity maps were generated for the two bands by running the \textsc{gMap} task, and were then used to select the source and background regions. A circle with radius of 7.4\,arcsec, which excluded most of the radiation from the host galaxy,  was selected as the source region for both NUV and FUV bands. An annulus with inner radius of 58\,arcsec and outer radius of 68\,arcsec was chosen as the background region for the FUV data. For the NUV data, a proper background region without contamination from a nearby source was not possible (only a concentered annulus background region is allowed in \textsc{gPhoton}). Thus background subtraction was not applied for the NUV data. We noted that this will not affect our conclusion on the NUV variability. The \textsc{gAperture} task was then used to calculate the AB magnitudes for both the NUV and FUV data.

\subsection{\mission{WISE}}
The Wide-field Infrared Survey Explorer \citep[WISE,][]{wright_etal2010} is a satellite that surveys the sky in mid-IR bands. We searched the ALLWISE \citep{mainzer_etal2011} and the NEOWISE Reactivation data release catalogues \citep{mainzer_etal2014} at the source position of \ngc with a matching radius of 1\,arcsec. This results in a total of 22 and 119 single exposure observations from the ALLWISE and NEOWISE, respectively. The W1 ($3.4\,\mathrm{\mu m}$) and W2 ($4.6\,\mathrm{\mu m}$) Vega magnitudes (contribution from the host galaxy was not subtracted) and uncertainties were obtained from the NASA/IPAC Infrared Science Archive (IRSA)\footnote{\url{https://irsa.ipac.caltech.edu}}.


\section{Results}

\begin{table*}
\begin{center}	
\caption{\label{tab:xray_fitting}Best-fitting parameters for the X-ray spectra.}
\hspace*{-1.5cm}
\begin{tabular}{cccccccc}\hline
 Observation ID & \nhost & $E_{\mathrm{Gaussian}}$/$T_{\mathrm{in}}$ & $\Gamma$ & $\log f_{\mathrm{0.5-10\,keV}}$ & $\log L_{\mathrm{0.5-10\,keV}}$ & $\mathrm{EW_{Gaussian}}$ & $C_{\mathrm{stat}}/\mathrm{d.o.f.}$ \\[1mm]
 & $10^{22}\,\mathrm{cm}^2$ & keV/eV &   & \unitflux & \unitlumi & eV & \\\hline
 &&&&\mission{XMM-Newton}&&& \\
 0066950301      & $<0.03$ & $6.46^{+0.10}_{-0.10}$  & $1.68^{+0.13}_{-0.08}$  &  $-12.03^{+0.04}_{-0.04}$  & $ 42.29^{+0.04}_{-0.04}$  & $523^{+619}_{-288}$  & 649.77/734    \\[1mm]
 0066950401      & $<0.05$ &          ---            & $1.73^{+0.17}_{-0.10}$  &  $-12.43^{+0.05}_{-0.06}$  & $ 41.89^{+0.05}_{-0.06}$  &     ---              & 599.15/696    \\[1mm]
 0305600601      & $<0.01$ & $6.43^{+0.05}_{-0.05}$  & $1.79^{+0.05}_{-0.05}$  &  $-11.58^{+0.02}_{-0.02}$  & $ 42.78^{+0.02}_{-0.02}$  & $523^{+362}_{-219}$  & 1086.11/1274  \\[1mm]
&&&&\mission{Swift/XRT}&&& \\
 00035365002$^a$ & $<0.05$ &          ---            & $1.31^{+0.48}_{-0.47}$  &  $-11.30^{+0.21}_{-0.21}$  & $ 43.02^{+0.21}_{-0.21}$  &    ---               & 46.74/24      \\[1mm]
 00035365002$^b$ & $<0.65$ & $19^{+15}_{-7}$         & $1.03^{+0.83}_{-0.77}$  &  $-11.21^{+0.24}_{-0.22}$  & $ 43.11^{+0.24}_{-0.22}$  &    ---               & 34.35/22      \\[1mm]
 2018 combined   & $<0.27$ &          ---            & $1.07^{+0.67}_{-0.53}$  &  $-12.98^{+0.24}_{-0.26}$  & $ 41.34^{+0.24}_{-0.26}$  &    ---               & 21.4/26       \\[0.5mm]\hline
\end{tabular}
\parbox[]{\textwidth}{
    $a$: Fitted with the baseline model.\\
    $b$: Fitted with the baseline model plus a blackbody component.
}
\end{center}
\end{table*}

\subsection{\label{sec:xray_var}X-ray variability}
\begin{figure*}
\begin{center}
\includegraphics[width=\textwidth]{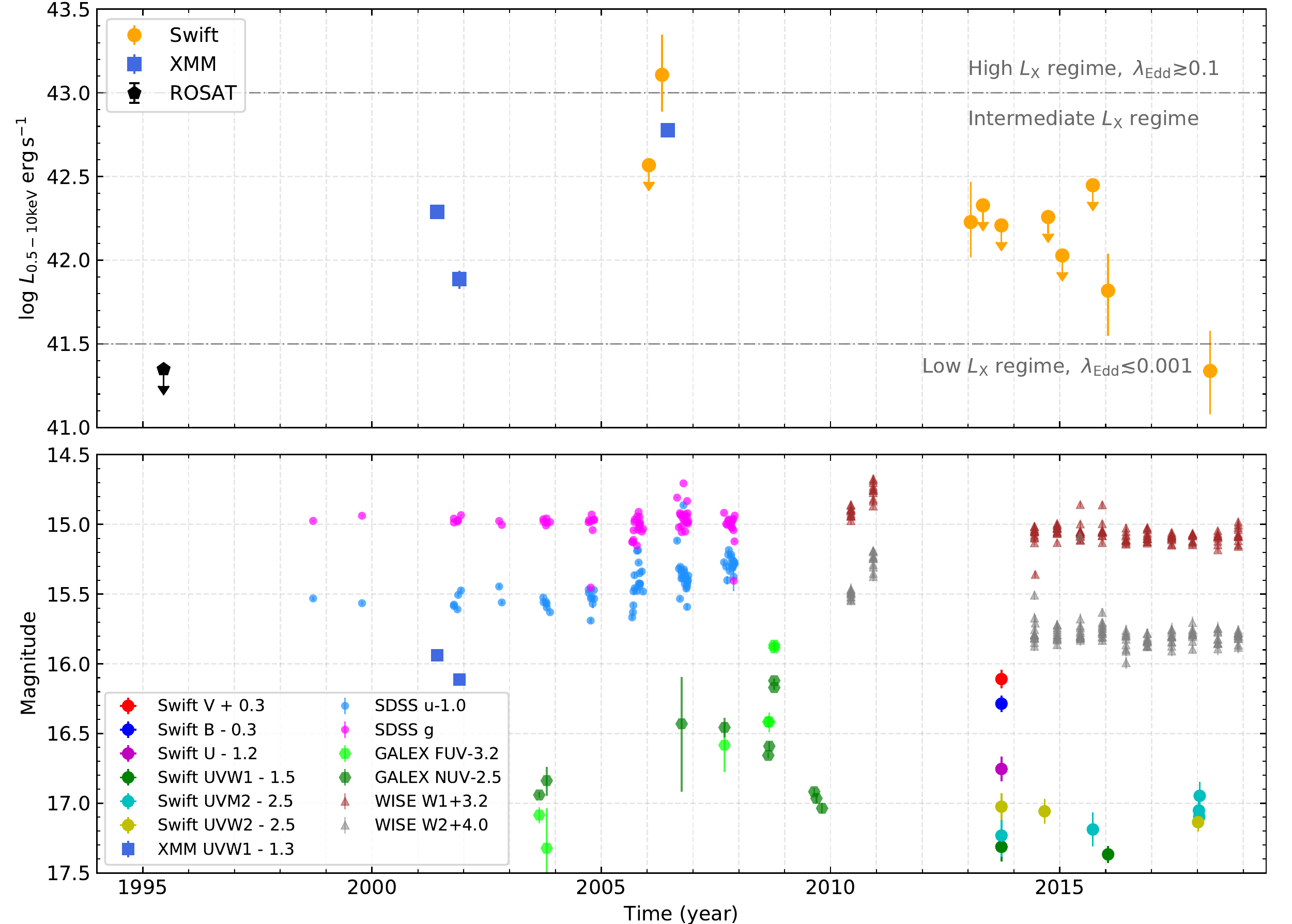}
\vspace*{-5mm}
\caption{\label{fig:multi_lc}Top panel: the intrinsic $0.5-10$\,keV band long-term X-ray light curve of \ngc. Black square: upper-limit estimated from 1995 \mission{ROSAT} observation; Royalblue triangles: the intrinsic flux was estimated by fitting the X-ray spectra of the three \mission{XMM--Newton} observations; Orange circles: the flux was estimated from X-ray spectra of \mission{XRT}, $3\sigma$ upper limits were given based on a power-law model for non-detections. The two dash-dotted grey lines mark the boundaries of the three luminosity regimes: high, intermediate, and low X-ray luminosity states, which are chosen as the \eddratio of being 0.1, 0.001, respectively, in this work. Bottom panel: the long-term optical (light blue circle for \textit{u} band, and magenta circle for \textit{g} band), \mission{GALEX/UV} (lime hexagon: FUV, green hexagon: NUV), \mission{Swift/UVOT} (filled circles, red:\textit{V}, blue: \textit{B}, magenta: \textit{U}, green: \textit{UVW1}, light green: \textit{UVM2}, olive: \textit{UVW2}), \mission{XMM--Newton/OM} (royalblue square: \textit{UVW1})}, and MIR (brown triangle for \textit{W1} and grey triangle for \textit{W2}) light curves for \ngc.
\end{center}
\end{figure*}

The long-term intrinsic 0.5--10\,keV flux ($\fluxfull$) variability of \ngc is shown in the upper panel of Fig.\,\ref{fig:multi_lc}. The $\fluxfull$ were estimated by fitting the X-ray spectra with an absorbed power-law model (see section \ref{sec:xray_spec}) for observations of which \ngc was detected. While the $3\sigma$ upper limits\footnote{The upper limit of $\fluxfull$ for the \mission{ROSAT} observation was roughly estimated by extrapolate the 0.1-2.4\,keV flux (see \citealt{yuan_etal2004}) assuming a power-law with photon index of 1.68, i.e. the best-fitting photon index of the 2001 June \mission{XMM--Newton} observation.}, calculated using the best-fitting parameters obtained by fitting the X-ray spectra of the nearest detection, were given for observations that the source was not detected.

From Fig.\,\ref{fig:multi_lc}, it is clear that \ngc showed large amplitude X-ray variability on a time-scale of $\sim10$ years. For instance, the highest observed X-ray luminosity (2006 April by \mission{Swift/XRT}) is more than 50 times higher than the lowest X-ray luminosity observed in 1995 by \mission{ROSAT} and 2018 by \mission{Swift/XRT}. Flare-like variations, over a time-scale of around half a year, were also found, e.g. one at around 2001 June and the other one at 2006 April. During those flaring phases, the X-ray luminosity changed by a factor of $\gtrsim3$. We note that flaring-like X-ray variability with time-scale of a few month may have also occurred in \ngc, as suggested from the multi-band optical/UV data (see Section\,\ref{sec:multband_lc}).

The fractional rms variability amplitude $F_\mathrm{var}$ \citep[e.g.][]{edelson_etal2002} is used to estimate the short-term (minutes to hours) variability. It is given by:
$$F_\mathrm{var} = \sqrt{\frac{S^2-\overline{\sigma^2_{\mathrm{err}}}}{\bar{x}^2}},$$
where $\bar{x}$ is the mean counts rate, $\overline{\sigma^2_{\mathrm{err}}}$ is the mean of the square of the $1\sigma$ errors, $S^2 = \sum_{i=0}^{N}(x_i-\bar{x})/(N-1)$ is the variance of the counts rate.

We calculated the fractional rms variability amplitude for the three \mission{XMM--Newton} observations. No significant short-term variability is found in all the three observations, i.e. with $F_\mathrm{var}$ less than a few per cent.

\subsection{\label{sec:xray_spec}X-ray spectral analysis}

\begin{figure}
\begin{center}
\includegraphics[width=\columnwidth]{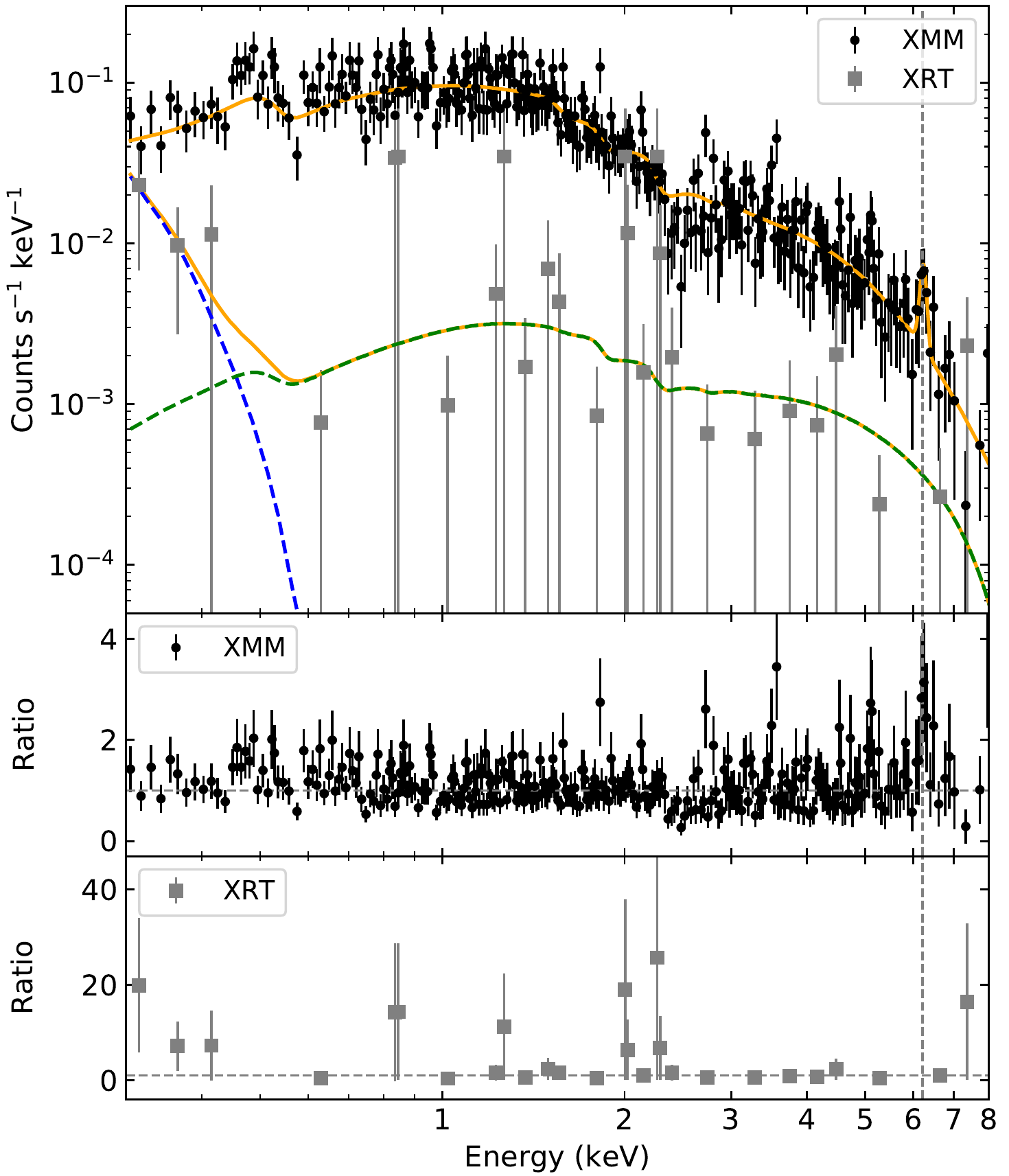}
\caption{\label{fig:xmm_xrt_spec} Top panel: the X-ray spectra of the 2006 \mission{XMM--Newton} (black circle, note that the combined MOS spectrum is shown in this plot) and the 2006 April \mission{Swift/XRT} observations (grey square). The best-fitting models for the two observations are shown as orange solid lines. The green dashed line represents the absorbed power-law continuum, while the blackbody component is shown as blue dashed line. The ratios of the data to the best-fitting continuum for the two observation are shown in the middle (\mission{XMM--Newton}) and bottom panels (\mission{Swift/XRT}). The vertical dashed line marks the 6.4\,keV Fe K$\alpha$ line in the observed frame.} 
\end{center}
\end{figure}

The X-ray spectra were analysed using \textsc{Xspec} \citep[version 12.10,][]{arnaud1996} with the Cash statistic (\citealt{cash1979}, wstat in \textsc{Xspec}). The energy band above $8$\,keV is dominated by background, we thus performed X-ray spectral modelling in the $0.3-8.0$\,keV energy range for all the data. For each of the \mission{XMM-Newton} observations, we jointly fitted all the spectra available from the EPIC cameras. A normalisation factor was thus added to account for the calibration differences between the instruments. Galactic and host galaxy absorptions were included in all our analysis using the \textsc{TBabs} and \textsc{zTBabs} model \citep[][abundances are set to wilm in \textsc{Xspec}]{wilms_etal2000}. The column density of the Galactic absorption was fixed at $3.84\times10^{20}\,\mathrm{cm}^{-2}$ \citep{kalberla_etal2005}. We defined an absorbed power-law model (\textsc{TBabs*zTBabs*zpo} in \textsc{Xspec}) as our baseline model. In the following spectral analysis, we define three different luminosity states (see the upper panel of Fig.\,\ref{fig:multi_lc}): the high luminosity state ($L_{0.5-10\,\mathrm{keV}}>10^{43}$\unitlumi), the intermediate luminosity state ($10^{41.5}$\unitlumi$<L_{0.5-10\,\mathrm{keV}}<10^{43}$\unitlumi), and the low luminosity state ($L_{0.5-10\,\mathrm{keV}}<10^{41.5}$\unitlumi). The two dot-dashed grey lines in the upper panel of Fig.\,\ref{fig:multi_lc} mark the boundaries of the three luminosity regimes. The Eddington ratios are calculated from the 2--10\,keV X-ray luminosity with a bolometric correction of 28 for the high and intermediate luminosity state, and 16 for the low luminosity state \citep{ho2008}.

\subsubsection{\label{sec:xrayhig_spec}High luminosity state: the 2006 April \mission{Swift/XRT} observation}
\mission{Swift/XRT} observed \ngc in its highest X-ray flux state observed so far on 2006 April 30 (one snapshot in observation 00035365002). The signal-to-noise ratio (S/N) of the X-ray spectrum is low (total background subtracted photon counts: 29) due to the short exposure time (412\,second). The best-fitting photon index is $1.31^{+0.48}_{-0.47}$ when fitted with the baseline model. An upper limit of $5\times10^{20}\,\mathrm{cm}^{-2}$ is estimated for the column density of the host galaxy. A significant excess in the soft X-ray band is shown in the ratio of data to the best-fitting baseline model (see the upper and bottom panels of Fig.\,\ref{fig:xmm_xrt_spec}). We thus fitted the spectrum with the baseline model plus a blackbody component. The likelihood ratio test (\textsc{lrt}) method was adopted to estimate the significance of the blackbody component. To perform the \textsc{lrt}, we simulated 10000 spectra using the best-fitting baseline model (\textsc{fakeit} in \textsc{Xspec}). The change in $C_{\mathrm{stat}}$ was then calculated by fitting the simulated spectra using both the baseline model and the model consisting of the baseline model and a blackbody component. The \textsc{lrt} result suggested that the blackbody component is significantly detected at more than $3\sigma$ level (99.9 per cent). The best-fitting temperature of the blackbody component is $19^{+15}_{-7}$\,eV, while the best-fitting photon index is $1.03^{+0.83}_{-0.77}$. An upper limit of $6.5\times10^{21}\,\mathrm{cm}^{-2}$ can be given for the host galaxy absorption. The predicted temperature of the inner region of a thin disc, e.g. $3-5\,R_{\mathrm{S}}$, close to the innermost stable circular orbit (ISCO), is $\sim 18$\,eV for a non-spin BH with mass of $10^7$\msun and accretion rate of 0.1, consistent with the best-fitting temperature of the blackbody component. The spectrum of a thin disc can generally be described as a multicolour blackbody, rather than a simple blackbody. We thus tried to fit the \mission{XRT} spectrum with the \textsc{diskbb} model. The best-fitting temperature ($18^{+7}_{-9}$\,eV) is consistent with that found in the simple blackbody model, though the normalization of the \textsc{diskbb} model can not be well constrained due to the low S/N and the narrow energy band. Simultaneously multiband observation\footnote{No \mission{Swift/UVOT} observations during the high luminosity period in 2006 April.} in the future could give a better constraint on the parameters in the \textsc{diskbb} model.

For source with relatively low photon counts, the \mission{Swift/XRT} spectrum could be affected by residual hot pixels or bright earth, both of which affect the low energy end of the spectrum (Andy Beardmore, private communication). No hot pixels are reported by the \textsc{xrthotpix} task, suggesting that the spectrum is not affected by residual hot pixels. Evidence for bright earth is found in the 2006 April \mission{Swift/XRT} observation (Obs. ID: 00035365002). However, we note that \ngc was observed in only one snapshot of this observation during which no sign of significant contamination from the bright earth was shown in the \mission{Swift/XRT} image.

To further assess the potential contribution from bright earth to the \mission{Swift/XRT} spectrum, we extracted a \mission{Swift/XRT} spectrum with the grade 0 events only\footnote{At low energy range, genuine X-rays should be grade 0 events, so the grade 0 and grade 0--12 spectra should look identical (Andy Beardmore, private communication).}. Except for the lowest energy bin ($<0.32$\,keV), which is slightly weaker (but still consistent within uncertainty), no obviously difference was found comparing to the grade 0--12 spectrum. A soft X-ray component which can be modelled by a blackbody component with temperature consistent with the value obtained from the grade 0--12 spectrum, is still detected at $3\sigma$ confidence level. Those results indicate that the soft excess shown in the 2006 April \mission{Swift/XRT} spectrum is unlikely caused by the bright earth.

\subsubsection{Intermediate luminosity state: the 2001 and 2006 \mission{XMM-Newton} observations\label{sec:xraymed_spec}}
The X-ray spectra of the first and second \mission{XMM--Newton} observations can be well fitted with the baseline model. The best-fitting results are $\Gamma=1.68^{+0.13}_{-0.08}$ and $N_\mathrm{H, host}<3\times10^{20}\,\mathrm{cm}^{-2}$ for the first observation, $\Gamma=1.73^{+0.17}_{-0.10}$ and $N_\mathrm{H, host}<5\times10^{20}\,\mathrm{cm}^{-2}$ for the second observation, consistent with that reported in \citet{yuan_etal2004}. The unabsorbed 0.5--10\,keV X-ray flux estimated using the best-fitting baseline model for the first and second observations are $9.5\times10^{-13}$ and $3.5\times10^{-13}$\unitflux, respectively. Evidence for an emission line profile at around 6.4\,keV, which was not reported in \citet{yuan_etal2004} due to the use of binned data, was found in the first observation. We then fitted the X-ray spectra with a model consisting of the baseline model and a Gaussian component, i.e. \textsc{TBabs*zTBabs*(zpo+zgau)} in \textsc{Xspec}. The line width $\sigma$ of the Gaussian component was fixed at $1$\,eV, i.e. a narrow emission line. The best-fitting line energy of the emission line profile is $6.46^{+0.10}_{-0.10}$\,keV, consistent with the narrow neutral Fe K$\alpha$ line that has been ubiquitously found in the X-ray spectra of AGN \citep[e.g.][]{nandra_etal2007, deLaCallePerez_etal2010}. The equivalent width of this line is $EW=523^{+619}_{-288}$\,eV. The best-fitting results can be found in Table\,\ref{tab:xray_fitting}.

In 2006, \mission{XMM--Newton} observed \ngc in an even higher X-ray luminosity state comparing to the two \mission{XMM--Newton} observations carried out in 2001. An emission line profile peaked at around $6.4$\,keV is clearly shown in the 2006 \mission{XMM--Newton} spectra (see upper and middle panels of Fig.\,\ref{fig:xmm_xrt_spec}). We thus fitted the X-ray spectra with the baseline model plus a Gaussian component (line width fixed at 1\,eV). This model can fit the data well with the best-fitting photon index of $1.79^{+0.05}_{-0.05}$ and an upper limit of $1\times10^{20}\,\mathrm{cm}^{-2}$ for the host galaxy absorption. The line energy is in agreement with the neutral Fe K$\alpha$ line, i.e. $E=6.43^{+0.05}_{-0.05}$\,keV, and the EW of the line is $523^{+362}_{-219}$\,eV which is consistent with EW found in the the 2001 observation. The significance of this line is higher than $3\sigma$ as inferred from the \textsc{lrt} method. Leaving the line width as a free parameter does not improve the fitting, and the best-fitting line width is $\sigma<115$\,eV, suggesting a narrow neutral Fe K$\alpha$ line. The EWs of the narrow Fe K$\alpha$ line, though with large uncertainties, are higher than the typical value ($\sim100$\,eV, e.g. \citealt{shu_etal2010}) found in broad line Seyfert galaxies , which may indicate a high Fe abundance or a large covering factor of the torus.

A soft X-ray excess component is often detected in the X-ray spectrum of AGNs \citep[e.g.][]{arnaud_etal1985, gierlinski_done2004}. This component can normally be fitted with a blackbody with temperature of $\sim150$\,eV \citep[e.g.][]{page_etal2004, gierlinski_done2004}. The \textsc{lrt} method is used to test whether a soft X-ray component is shown in the 2001 and 2006 \mission{XMM--Newton} observations. We found that the soft X-ray component is not significantly (less than $2\sigma$) detected in the 2006 \mission{XMM--Newton} observation. Evidence for the soft X-ray excess was also not found in the 2001 \mission{XMM--Newton} observations.  

\subsubsection{Low luminosity state: the 2018 \mission{Swift/XRT} observations\label{sec:xraylow_spec}}
As mentioned in section\,\ref{sec:xray_var}, the lowest X-ray flux was detected in the 1995 \mission{ROSAT} and the combined 2018 \mission{Swift/XRT} observations. Hereafter, we refer those observations as the low luminosity state. The combined 2018 \mission{Swift/XRT} spectrum, with a total exposure time of $25$\,ks, can be fitted with the baseline model (see Table\,\ref{tab:xray_fitting}), though with large uncertainties due to low S/N (26 net source counts in the 0.3--8\,keV band). The column density of the host galaxy cannot be well constrained, with a 90 per cent upper limit of $2.7\times10^{21}\,\mathrm{cm}^{-2}$. The best-fitting photon index is rather flat but with large uncertainty, i.e. $\Gamma=1.07^{+0.67}_{-0.53}$, which is still consistent with the photon index estimated at the intermediate and high X-ray luminosity states (see Section\,\ref{sec:xraymed_spec} and \ref{sec:xrayhig_spec}).

\begin{table*}
\begin{center}
\caption{\label{tab:pcf_fitting}Best-fitting parameters for the partial covering model.}
\begin{tabular}{lccccccc}\hline
Model & Observation ID & $\Gamma$ & $\log f_{\mathrm{0.5-10\,keV}}$ & $N_{\mathrm{H}}$          & $f_\mathrm{cov}$ & $\log\xi$ & $C_{\mathrm{stat}}/\mathrm{d.o.f.}$ \\[1mm]
      &                &          &     \unitflux              & $10^{22}\,\mathrm{cm}^2$  &                  &           &                                     \\\hline
Partial covering         & 0305600601 & $1.87^{+0.08}_{-0.07}$ & $-11.46^{+0.16}_{-0.10}$ & $20^{+63}_{-15}$    & $0.30^{+0.20}_{-0.20}$ &                     & 1080.22/1273 \\[1mm]
                         & 0066950301 & $1.72^{+0.11}_{-0.10}$ & $-11.75^{+1.38}_{-0.24}$ & $58^{+362}_{-52}$   & $0.50^{+0.49}_{-0.20}$ &                     & 653.49/735   \\[1mm]
                         & 0066950401 & $1.86^{+0.31}_{-0.23}$ & $-12.42^{+0.07}_{-0.06}$ & $<0.4   $           & $<0.6        $         &                     & 598.37/695   \\[1mm]
                         & 0003456789 & $2.92^{+1.49}_{-1.33}$ & $-10.71^{+0.86}_{-0.55}$ & $5^{+6}_{-2}$       & $>0.54       $         &                     & 42.36/23    \\[1.5mm]
Ionized partial covering & 0305600601 & $1.75^{+0.06}_{-0.06}$ & $-11.50^{+0.27}_{-0.07}$ & $64^{+340}_{-60}$   & $0.64^{+0.36}_{-0.40}$ & $3.4^{+0.8}_{-1.3}$ & 2357.90/2712 \\[1mm]
                         & 0066950301 & $1.70^{+0.14}_{-0.12}$ & $-11.67^{+2.68}_{-0.34}$ & $63^{+366}_{-54}$   & $0.58^{+0.35}_{-0.50}$ & $<4.0 $        & 653.46/734   \\[1mm]
                         & 0066950401 & $1.63^{+0.12}_{-0.11}$ & $-12.33^{+0.41}_{-0.05}$ & $179^{+114}_{-174}$ & $>0.5       $          & $3.7^{+0.6}_{-1.4}$ & 595.87/694   \\[1mm]
                         & 0003456789 & $2.51^{+2.33}_{-0.99}$ & $-10.80^{+0.64}_{-0.46}$ & $5^{+7}_{-4}$       & $>0.86      $          & $<2.5$              & 39.99/22    \\[2.0mm]

\multicolumn{8}{c}{Joint fit the \mission{XMM--Newton} and \mission{Swift/XRT} observations} \\[1.5mm]
Partial covering         & 0305600601 & $1.77^{+0.05}_{-0.04}$ & $-11.27^{+0.78}_{-0.25}$ & $89^{+196}_{-74}$   & $0.50^{+0.41}_{-0.40}$ &                     & 2377.40/2736 \\[1mm]
                         & 0066950301 &    ---                 &    ---                   & $156^{+201}_{-78}$  & $0.84^{+0.13}_{-0.13}$ &                     &              \\[1mm]
                         & 0066950401 &    ---                 &    ---                   & $      >219    $    & $0.93^{+0.06}_{-0.05}$ &                     &              \\[1mm]
                         & 2018--XRT  &    ---                 &    ---                   & $      >107    $    & $0.99^{+0.01}_{-0.01}$ &                     &              \\[1.5mm]
Ionized partial covering & 0305600601 & $1.72^{+0.04}_{-0.04}$ & $-11.12^{+0.29}_{-0.36}$ & $165^{+141}_{-16}$  & $0.79^{+0.14}_{-0.20}$ & $2.9^{+0.1}_{-0.5}$ & 2368.01/2732 \\[1mm]
                         & 0066950301 &    ---                 &    ---                   & $148^{+180}_{-82}$  & $0.88^{+0.06}_{-0.09}$ & $<2.6             $ & 	      \\[1mm]
                         & 0066950401 &    ---                 &    ---                   & $      >56     $    & $0.95^{+0.03}_{-0.08}$ & $<1.9      $        & 	      \\[1mm]
                         & 2018--XRT  &    ---                 &    ---                   & $      >87     $    & $0.99^{+0.01}_{-0.01}$ & $<2.7      $        &              \\\hline
\end{tabular}
\end{center}
\end{table*}

\subsubsection{Alternative model: partial covering}
\citet{yuan_etal2004} found that the X-ray spectra of the two 2001 \mission{XMM--Newton} observations can also be well fitted with a partial covering model (see their table 2). We also tried to fit the \mission{XMM--Newton} and \mission{Swift/XRT} data with the partial covering model, i.e. \textsc{TBabs*TBpcf*zpo} in \textsc{xspec} (or \textsc{TBabs*zxipcf*zpo} for ionized partial covering model).

We first independently fitted the three \mission{XMM--Newton} and the 2006 April \mission{XRT} spectra. Both the column density and the covering factor parameters cannot be constrained for the 2018 \mission{XRT} spectrum, thus this observation was not fitted with the partial covering model. We found that the (ionized) partial covering model can fit the X-ray spectra. However, it does not improve the fitting significantly comparing to the simple baseline model and some of the parameters cannot be well constrained (only upper/lower limits can be given, see Table\,\ref{tab:pcf_fitting}). 

If indeed the observed X-ray variability was due to partial covering absorption, then we may expect that the intrinsic primary emission did not change significantly throughout the years covering all the X-ray observations. We thus fitted the three \mission{XMM--Newton} and the combined 2018 \mission{Swift/XRT} observations\footnote{The parameters cannot be constrained for the 2006 \mission{XRT} data when jointly fitted with all the other X-ray spectra. We thus did not include this observation in our joint X-ray spectral analysis.} simultaneously with the partial covering model. The parameters in the power-law model, i.e. the photon index $\Gamma$ and the normalization, were linked for the X-ray spectra of the four observations, while the column density and the covering factor were fitted independently for each individual observation. The best-fitting results of the joint spectral analysis can be found in Table\,\ref{tab:pcf_fitting}. The (ionized) partial covering model can simultaneously fit the X-ray data well. However, the inferred covering factors are relatively large, especially for the 2018 \mission{Swift/XRT} observation of which a covering factor close to 1 is required.

\subsection{\label{sec:multband_lc}Variability in optical, UV and mid-infrared}
In the bottom panel of Fig.\,\ref{fig:multi_lc}, we also show the long-term light curves in optical (light blue: \mission{SDSS} $u$ band, magenta: \mission{SDSS} $g$ bands), UV (\mission{GALEX} data, lime and green hexagon for FUV and NUV, respectively), and mid-infrared (WISE, brown circle: W1, gray circle: W2). Large variability with $\Delta{m}>0.5$\,mag is clearly seen in the mid-infrared to optical/UV bands.

The long-term SDSS $u$ and $g$ bands light curves are shown in the lower panel of Fig.\,\ref{fig:multi_lc}. The light curve in the $r$, $i$, and $z$ bands are similar to the $g$ band but with smaller amplitude of variability. From Fig.\,\ref{fig:multi_lc}, it is suggested that \ngc was in a relatively low flux state before 2005. A `flaring-like' profile, with a duration likely longer than 3 months, can be seen in both the $u$ and $g$ bands during 2005-2006. The source went into an intermediate flux state in 2006-2008 as inferred from the SDSS data.

The characteristic of the NUV and FUV data are consistent with the optical light curve. As in the optical, the NUV and FUV magnitudes of \ngc were relatively low in the 2003 observation, while they increased to an intermediate level during 2006-2008. In addition, a rapid rise (probably a flare) within 1.3 months is clearly shown in the 2009 observations. The source went into a low flux state in NUV in 2009. Weak variability ($\Delta m < 0.2\,$mag) are also shown in the UV light curves measured from the \mission{Swift/UVOT} and \mission{XMM/OM}. We caution here that such a small magnitude of variability may caused by systematic uncertainties in the aperture photometric measurements, especially for the \mission{Swift/UVOT} measurements of which the host galaxy may dominate the UV radiation as the source was in a very low luminosity state during the observations.

The \mission{WISE} W1 and W2 mid-infrared magnitude measured at around the end of 2010 is about 0.2\,mag higher than that observed half a year earlier. No significant variability is seen in the data taken after 2014, which are $\sim0.5$\,mag dimmer than previous observations. Those data suggest that, though the sampling is not good due to the lack of data from 2011 to 2014, a `flaring' like profile is probably also shown in the \mission{WISE} W1 and W2 data. The $W1$ and $W2$ magnitudes did not vary significantly after 2014, which may be due to the fact that the radiation in the two bands were dominated by the host galaxy.

The `averaged' luminosity is $1.5\times10^{41}$\unitlumi in 0.5--10\,keV band, as estimated from the measured narrow H$\alpha$ luminosity using the H$\alpha$--X-ray luminosity relation \citep{yuan_etal2004}. The long-term multi-band light curves indicate that probably \ngc normally behaves like a LLAGN, and underwent several outbursts during the last three decades.

\section{Discussion}
Using only the 1995 \mission{ROSAT} and 2001 \mission{XMM--Newton} observations, \citet{yuan_etal2004} suggested that the large X-ray amplitude, with a change in X-ray flux by a factor of $>10$, can be explained by intrinsic variability, i.e. the change of accretion rate. However, a partial covering mode can also explain the observed variability, and cannot be ruled out. In this work, using more X-ray data including new observations from \mission{XMM--Newton} and \mission{Swift/XRT}, we found that the \ngc can reach an even higher X-ray luminosity, resulting in an increase in X-ray luminosity by a factor of more than 50 comparing to the 1995 \mission{ROSAT} and the 2018 \mission{Swift/XRT} observations. In addition, flare-like features are also shown in the optical, UV, and MIR long-term light curves, suggesting that \ngc underwent several `outburst' in the past decades.

\subsection{Intrinsic variability vs partial covering absorption}
Those new X-ray observations and multi-band data allowed us to further test the (ionized) partial covering scenario. In this scenario, we assumed that the flux of the primary emission was the same during different X-ray observations. We then jointly fitted the 2018 \mission{Swift/XRT} observations (low luminosity state) with the 2001 and 2006 \mission{XMM--Newton} observations (intermediate luminosity state). The (ionized) partial covering model can fit the X-ray data well (see Table\,\ref{tab:pcf_fitting}). However, the inferred covering factors are relatively large, especially for the 2018 \mission{Swift/XRT} observation of which a covering factor close to 1 is required. The result indicates that the primary emission was nearly fully obscured by an approximately Compton-thick absorber during the 2018 \mission{Swift/XRT} observation. However, this is at odds with the best-fitting results obtained by fitting the X-ray spectrum of the combined 2018 \mission{XRT} observations with the baseline model of which no evidence for heavily absorption is found, indicating that the primary emission is unlikely to be constant.

Further evidence against the partial covering model can be found from the long-term MIR variability. As mentioned before, \ngc showed flare-like features in MIR bands (see Fig.\,\ref{fig:multi_lc} and Section\,\ref{sec:multband_lc}). The MIR emission in AGN is believed to originate from the reprocessing of the primary emission (mainly the UV/optical radiation\footnote{A positive correlation between the UV and X-ray flux is found in some AGNs during the flare phase, e.g. NGC 1556 \citep{oknyansky_etal2019} and Mrk 335 \citep{gallo_etal2018}. Thus a flare in the X-ray band may also imply a burst in the UV/optical band.}) by the dusty torus. In the partial covering scenario, the covering factor along the line-of-sight (LOS) can change dramatically (thus cause the X-ray variability). However, the global covering factor of the absorber (as well as the dusty torus) is not expected to vary significantly. So the fraction of the primary emission that will be reprocessed and re-radiated in the MIR band by the dusty torus should not show large amplitude variability in a relatively short time-scale. The significant variability found in the MIR data of \ngc is then not expected in the partial covering scenario. Thus the large amplitude X-ray variability in \ngc is unlikely caused by partial absorption.

As mentioned in Section\,\ref{sec:multband_lc}, `flare-like' features are also seen in the long-term UV/optical light curve. The sudden increase of the MIR flux in around 2011 was more likely to be attributed to the echo of a burst in the UV/optical band. Evidence for MIR echos had been found in the changing-look AGNs \citep[e.g.][]{sheng_etal2017}, and in the TDEs \citep[e.g.][]{dou_etal2016, dou_etal2017, jiang_etal2016, jiang_etal2017}. We conclude that the observed multi-band variabilities were caused by the change of the intrinsic radiation rather than partial covering absorption.

\subsection{Accretion mode transition in \ngc?}
Theoretical calculation suggests that a transition in accretion mode will occur once the accretion rate reaches the critical values \citep[][]{meyer_etal2000}. Such a transition has been observed in Galactic XRBs \citep{tanaka_etal1995, zhang_etal1997}. The predicted time-scale for the accretion mode transition in AGN is much longer than that in XRBs, making it difficult to find such a transition system among AGNs. Conclusive evidence for accretion mode transition in AGN has not been found yet. 

\ngc is spectroscopically classified as Seyfert 1.9 or LINERS from the 2000 SDSS spectrum (see \citealt{yuan_etal2004}). In the above subsection, we argue that the large amplitude X-ray variability was caused by the change of intrinsic radiation rather than absorption. The unabsorbed intrinsic 2--10\,keV X-ray luminosity in the low luminosity state is then well below $\sim10^{42}$\,\unitlumi, i.e. in the LLAGN regime. \citet{yuan_etal2004} proposed that the accretion proceeded via RIAF/ADAF during the low X-ray luminosity state observed by \mission{ROSAT} in 1995, and that the large X-ray amplitude, with an increase by a factor of $>10$ from the 1995 \mission{ROSAT} to the 2001 \mission{XMM--Newton} observations, can be explained by a transition in accretion mode, i.e. from a RIAF/ADAF to a standard thin disc or vice verse. During the 2001 \mission{XMM--Newton} observations, the accretion flow in \ngc was possibly in a transition state between ADAF and thin disc. \citet{yuan_etal2004} also predicted that a transition to thin disc might have taken place if the flux could reach a much higher peak value than the 2001 \mission{XMM--Newton} observations. In this work, using latest X-ray data from \mission{XMM--Newton} and \mission{Swift/XRT}, we found that \ngc can indeed reach an even higher X-ray flux, resulting in an increase in X-ray flux by a factor of more than 50.

\ngc was in a high X-ray luminosity state during the 2006 April \mission{Swift/XRT} observation, with an estimated \eddratio of $\sim0.13$ (assuming a X-ray bolometric correction factor of 28, \citealt{ho2008}). The estimated \eddratio, which is $>100$ times the \eddratio found in the 1995 low luminosity state, is well beyond the critical values predicted by theoretical calculation, indicating a thin disc mode in \ngc during the 2006 April observation. Evidence for the emergence of a thin disc is further supported from the X-ray spectral analysis. A soft X-ray component which can be fitted with a blackbody model is significantly detected in the 2006 April \mission{Swift/XRT} spectrum. Unlike the soft X-ray excess (normally has a temperature of $\sim150$\,eV when modelled with a blackbody) that commonly found in Seyfert 1 galaxies, the best-fitting temperature of this component, which is $\sim19^{+15}_{-7}\,\mathrm{eV}$, is in agreement with the expected temperature of the inner region of a thin disc with BH mass of $\sim10^7$\msun. Such a component has been reported for only a few sources so far, e.g. RXJ1643+49 \citep[][see also \citealt{shu_etal2017, sun_etal2013} for the other sources]{yuan_etal2010}. We conclude that a transition from ADAF to the thin disc had taken place in the 2006 April observation.

\ngc was observed to exhibit a low X-ray luminosity state, i.e. comparable to the 1995 \mission{ROSAT} observation, during the 2018 \mission{Swift/XRT} observations. The \eddratio is conservatively estimated to be $\sim10^{-3}$ (assuming a X-ray bolometric correction factor of 16, \citealt{ho2008}) using the 2--10\,keV luminosity measured from the combined 2018 \mission{Swift/XRT} observation. This may indicate an accretion flow via ADAF, suggesting a transition from the thin disc to ADAF. More evidences to support the accretion mode transition, such as the broad band SED in the low luminosity state, the reflection spectrum, and the broad Fe K$\alpha$ line, might be found in future X-ray and multi-band observations with better data quality.

Our results may imply that accretion mode transition can indeed take place in AGN, supporting the idea proposed by \citet{yuan_etal2004}. It also proves the idea that the accretion theory can apply to BH--accretion system across different BH mass scales. Future simultaneously multi-band observations of \ngc, especially optical spectroscopic data, will potentially help us understanding some important but yet unclear questions in AGN, such as the long-term variability of AGN \citep[e.g][]{cao_wang2014} and the changing-look AGN.

\subsection{Transition time-scale}\label{subsec:trans_timescale}
The physical mechanism of the state transition from the low state to the high state (ADAF to thin disc) or vice versa (thin disc to ADAF) could be understood in the framework of the disc evaporation model \citep[e.g.][]{liu_etal1999, meyer_etal2000, meyer_etal2000b, liu_etal2009, taam_etal2012, qiao_etal2013}. The disc evaporation model predicts an evaporation-curve, in which the evaporation rate increases with decreasing the radius until a maximum evaporation rate $\dot M_{\mathrm{max}}$ is reached and then the evaporation rate decreases with decreasing the radius. So in the disc evaporation model, if the mass accretion rate from the outer region of the disc is less than $\dot M_{\mathrm{max}}$ ($\sim$ a few percent of the Eddington accretion rate $\dot M_{\mathrm{Edd}}=L_{\mathrm{Edd}}/c^2$, where $L_{\mathrm{Edd}}=1.3 \times 10^{38}\,M_{\mathrm{BH}} / M_{\odot}$\,\unitlumi), the disc will be truncated at a radius where the mass accretion rate equals the evaporation rate. From the truncation radius inwards, the accretion flow will be existed in the form of the ADAF. Then the geometry of the accretion flow will be an inner ADAF plus an outer truncated disc. Such a geometry of the accretion flow is often used to explain the spectral features of the low state of black hole X-ray binaries \citep[e.g.][]{esin_etal1997} and low-luminosity AGNs \citep{ho2008, nemmen_etal2011}. Whereas, if the mass accretion rate from the outer region of the disc is greater than $\dot M_{\mathrm{max}}$, the disc cannot be completely evaporated at any radius, and the disc will extend down the ISCO of the black hole. Such a geometry of the accretion flow is widely used to explain the spectral features of the high state of black hole X-ray binaries and luminous AGNs.

The time-scale of the state transitions might be helpful in understanding the physical process (e.g. the disc evaporation model) for accretion mode transition. Assuming that the flare-like feature showed in the multi-band data were due to the change of intrinsic radiation, and a transition of accretion had taken place, we can then roughly estimate the transition time-scales:\\
\textbf{Low state to high state:} the time-scale for the transition from low state to the high state, or from ADAF to thin disc (hereafter, $t_\mathrm{ADAF\rightarrow TD}$), estimated from the X-ray data alone is likely to be several months to less than 4 years, as suggested from the 2001 and 2006 X-ray observations. More stringent constraints on this transition time-scale could be obtained from the optical and UV data. As can be seen from Fig.\,\ref{fig:multi_lc}, the variability time-scale is likely to be longer than $\sim2$ months and shorter than $\sim1$ year estimated from the optical (the raising phase in 2005) and NUV/FUV (the raising phase of the UV flare in 2008) data. However, the relatively small amplitude variability ($\sim0.5$\, mag) of those flares, comparing to the X-ray variability and the decline phase of the 2008 NUV flare ($\sim1$\,mag), may indicate that the observations did not cover the whole transition process, and/or the transition was not complete. The transition time-scale $t_\mathrm{ADAF\rightarrow TD}$ is then roughly estimated to be a few years ($<4$\,years) by combing the X-ray and optical/UV data.\\[1mm]
\textbf{High state to low state:} a time-scale of $\sim12$\,years for the transition from high state to low state, or from thin disc to ADAF (hereafter, $t_\mathrm{TD\rightarrow ADAF}$), can be inferred from the long-term X-ray variability observed by \mission{Swift/XRT} from 2006--2018. However, a much shorter time-scale can be inferred from the NUV data. From the bottom panel of Fig.\,\ref{fig:multi_lc}, it is clear that the NUV flux peaked at around late 2008, then it gradually declined to a low flux state (comparable to the lowest NUV flux detected in late 2003) till late 2009, probable implying a transition to the ADAF accretion flow. If this is the case, then the transition time-scale $t_\mathrm{TD\rightarrow ADAF}$ could be as short as $\sim1$\,year.

\subsubsection{Transition from low state to high state}
In the present work, initially when the source is in the low state, the disc is suggested to be truncated at a radius of a few hundreds of Schwarzschild radii as predicted by the disc evaporation model \citep{taam_etal2012}. The disc component detected in the 2006 April \mission{Swift/XRT} observation suggests that the inner region of the accretion disc extends probably to the ISCO. The time-scale of the transition from the low state to the high state may correspond to the viscous time-scale of the truncated disc extending down to the ISCO of the black hole as the mass accretion rate in the disc exceeding $\dot M_{\mathrm{max}}$. Such a viscous time-scale $\tau_\mathrm{vis}$ can generally be calculated as $\tau_{\mathrm{vis}} \sim R / v_{R}=R^{2} /\left(\alpha c_{\mathrm{s}} H\right)$ \citep{frank_jhan_king2002, kato_fukue_mineshige2008}, where $\alpha$ is the poorly known viscosity parameter; $H$ is the scale height of the accretion disc; $v_{R}$ and $c_{\mathrm{s}}$ are the radial velocity and the sound speed at radius $R$, respectively. Following \citet{cao_wang2014}, we estimate the viscosity time-scale at a given truncation radius $R_{\mathrm{tr}}$ using the following equation:
\begin{equation}
\begin{aligned}
\label{eq:vis}
\tau_{\mathrm{vis}} & \sim \frac{R_{\mathrm{tr}}}{v_{R}}=\frac{R_{\mathrm{tr}}^{2}}{\alpha c_{\mathrm{s}} H} \\
&=1.56 \times 10^{-7} r_{\mathrm{tr}}^{3 / 2} \alpha^{-1}\left(\frac{H}{R}\right)^{-2}\left(\frac{M_{\mathrm{BH}}}{10^{6} \mathrm{M}_{\odot}}\right) \mathrm{yr},
\end{aligned}
\end{equation}
where $r_{\mathrm{tr}}=R_{\mathrm{tr}}/R_{\mathrm{g}}$ is the inner truncation radius (in units of gravitational radius) of the accretion disc, while $H/R$ is the disc height to radius ratio which is normally less than 0.1 for a thin disc. By modeling the SED of a sample of LLAGNs using an accretion-jet model, \citet{nemmen_etal2011} suggested that the thin disk in LLAGNs is truncated at $60-450\,R_{\mathrm{g}}$. Assuming the disc is truncated at a few hundreds $R_{\mathrm{g}}$ (e.g. $150\,R_{\mathrm{g}}$) in \ngc, a $H/R\sim0.05$ is required so that the predicted viscosity time-scale (4 years for a disc with $\alpha=0.5$) matches with the estimated $t_\mathrm{ADAF\rightarrow TD}$ (a few years) for \ngc. Although such a large $H/R$ is not expected in the thin disc model ($H/R$ should be $\sim10^{-3}$), it can be obtained in the magnetic pressure supported accretion disc model proposed by \citep{dexter_begelman2019}. In this model, the disc is geometrically thick at all luminosities which gives a viscous propagation time-scale as short as a few years, consistent with the observed time-scale.

Alternatively, the thermal-viscous accretion disc instability model (DIM) \citep{meyer_meyer1981}, which is caused by the ionization of the hydrogen at the outer disc, has been successfully explained the transient outbursts observed in X-ray binaries and cataclysmic variables (CVs). DIM is proposed in \citep{noda_done2018} as one possible explanation for the luminosity change in Mrk 1018. In the DIM scenario, the jump of the temperature of the outer disc, triggered by ionization disc instability when the outer disc temperature is high enough to ionize hydrogen, will eventually results in a heating front propagating inward, leading to an increase in flux/luminosity. The time-scale is determined by the travelling speed of this heating front, which is shorter than the viscous time-scale by a factor of $H/R$. However, as noted by \citep{noda_done2018} (see also the simulations in \citealt{hameury_etal2009}), the time-scale will still be too long to match the observed time-scale in \ngc and Mrk 1018. A large $H/R$ is still required to explain the observed time-scale.

\subsubsection{Transition from high state to low state} 
The transitional time-scale from the high state to the low state can be estimated in the framework of the disc evaporation model as the mass accretion rate decreasing just below $\dot M_{\mathrm{max}}$. When the mass accretion rate is just below $\dot M_{\mathrm{max}}$, the disc will first be truncated at a critical radius $R_{\mathrm{crit}}$ corresponding to $\dot M_{\mathrm{max}}$, leaving an inner disc and an outer disc at the same time. The inner disc will be completely swallowed into the black hole within the viscous time-scale. Such a viscous time-scale (hereafter $\tau_{\mathrm{dic}}$) is related with the critical truncation radius of the disc $R_{\mathrm{crit}}$, which can be from a few tens to a few hundreds of Schwarzschild radii depending on the viscosity parameter $\alpha$, and the strength of the magnetic field \citep{qian_etal2007, taam_etal2012}. The estimated $\tau_{\mathrm{disc}}$ could be as short as $\sim2$ months for $R_{\mathrm{crit}}=20\,R_{\mathrm{g}}$, assuming $\alpha=0.3$ and $H/R=0.05$ (again, such a large $H/R$ can be obtained in the magnetic pressure supported disc model). Observationally, the disc component shown in the 2006 April \mission{Swift/XRT} observation was not significantly detected (less than $2\sigma$) in the \mission{XMM--Newton} observation carried out one and half month later. This time-scale is consistent with the estimated $\tau_{\mathrm{disc}}$ for $R_{\mathrm{crit}}=20\,R_{\mathrm{g}}$, indicating a small critical truncation radius at $\dot M_{\mathrm{max}}$ for \ngc. Here we note that the $\alpha$ and $H/R$ parameters can also have a large impact on the viscosity time-scale. A more precisely measurement of the transition time-scale in the future may be helpful to roughly constrain these parameters.

On the other hand, the inner radius of the outer disc ($R_{\mathrm{in, outer}}$) may move towards to a much larger radii from $R_{\mathrm{crit}}$ as the mass accretion rate decreases, providing that the evaporation rate is higher than the mass accretion rate at $R_{\mathrm{in, outer}}$ of the outer disc. Eventually the outer disc will truncate at a radii ($R_{\mathrm{tr, outer}}$) where the evaporation rate equals to the mass accretion rate. If this is the case, then the transition time-scale depends on not only the evaporation rate but also the decline curve of the mass accretion rate, both of which are subject to large uncertainties. Thus it is difficult to give a quantitate estimation of the high state to low state transition time-scale from the disc evaporation theory based on the current data of \ngc. Nevertheless, we may be able to test the disc evaporation model and constrain some of the key parameters in the model with better sampled multi-band light curves and more precisely measured transition time-scale in the future observations.

In the DIM scenario, the decrease in luminosity/flux can be explained as the propagation of the cool front, which is triggered when the temperature of the outer disc drops below $10^4\,$K, i.e. the ionized disc becomes neutral. Again, similar to the time-scale for the propagation of the heating front, a large $H/R$ is required to explain the observed decline time-scale.

\subsection{The soft X-ray excess}
The soft X-ray excess has been commonly found in the X-ray spectra of AGNs. However, its origin is still unclear. The temperature of this component, when modelled with thermal Comptonization model, is $\sim0.12$\,keV \citep{gierlinski_done2004} which is much higher than the predicted temperature of the accretion disc with BH mass $>10^6$\,\msun but is consistent with thermal Comptonization of a warm corona \citep{magdziarz_etal1998}. Evidence for a soft X-ray component originated from a thin accretion disc, however, has been found in a few AGNs \citep{yuan_etal2010, sun_etal2013, shu_etal2017}. The temperature of the soft X-ray excess in \ngc is $\sim19$\,eV (modelled with a blackbody) in the high luminosity state (with \eddratio higher than 0.1), suggesting that this component was from the accretion disc during the 2006 April \mission{Swift/XRT} observation. However, no significantly soft X-ray excess is detected in the 2001 (June and November) and 2006 (June) \mission{XMM--Newton} observations during which the source was in an intermediate luminosity state, implying that the soft X-ray excess component could disappear within two months in \ngc.

The property of the soft X-ray excess for \ngc is very similar to the results reported in \citet{noda_done2018} for the changing-look AGN Mrk 1018. The soft X-ray excess component in Mrk 1018 also emerged in the bright state ($\lambda_{\mathrm{Edd}}>0.02$), and it disappeared in the faint state ($\lambda_{\mathrm{Edd}}>0.02$). Unlike \ngc, the temperature of the soft X-ray component in the bright state in Mrk 1018 is consistent with the Comptonization of a warm corona. While the emergence and disappearing of the soft X-ray component in Mrk 1018 is explained as the increase and decrease of the size of the warm corona region \citep{noda_done2018}, respectively, the non-detection of the soft X-ray component in \ngc in the intermediate state can be explained by the truncation of the inner accretion disc under the framework of the disc evaporation as discussed in Section\,\ref{subsec:trans_timescale}. Whether a warm corona has ever formed in \ngc is unknown due to the lack of observational data.

It is worth noting that the property of the soft X-ray excess also changed within two months for Mrk 1018 as reported in \citep{noda_done2018}. These results indicate that the soft X-ray excess in AGNs can be a transient phenomenon. The most impressive evidence for the transition/variability of the soft X-ray excess component came from the recently discovered quasi periodic eruptions in GSN 069 \citep[][see also RX J1301.9+2747, \citealt{sun_etal2013, miniutti_etal2019}]{miniutti_etal2019}. During the late 2018 to early 2019 X-ray observations, a cold soft X-ray component with temperature of $\sim50$\,eV is shown in the quiescent state of GSN 069, while a warm soft X-ray component with temperature of $\sim120$\,eV is found at the peak of the eruptions which last for several kilo-seconds with an recurrence time of $\sim30$\,ks. Moreover, the cold component can increase its temperature to $\sim80$\,eV within one month. We note that the physical mechanism that triggers the emergence of the warm component and the transition between the cold and warm components may be very different in GSN 069, as the time-scale is much shorter than that observed in Mrk 1018 and \ngc. Nevertheless, these results indicate that there may be a diversity in the origin of the soft X-ray excess. A detail temporal analysis of the soft X-ray excess, on both short and long time-scales, may provide important insights on the origin of the soft X-ray excess and the mechanism triggering the accretion mode transition.

\section{Summary}
In this paper, we report the discovery of large amplitude X-ray variability, i.e. by a factor of more than 50, of the LLAGN \ngc using the archival X-ray data span several decades. The long-term X-ray, optical, and UV variability (see Fig.\,\ref{fig:multi_lc}) indicate that \ngc may have undergone several outbursts in the past decades. A detail analysis of the X-ray spectra in different X-ray luminosity state, combining with the MIR variability, suggests that the large amplitude X-ray variability is likely caused by the change of the intrinsic X-ray flux, rather than due to the (partial) absorption. 

Possible evidence for accretion mode transition in \ngc is found by modelling the X-ray spectra at different luminosity state. At low X-ray luminosity state (during the 2015 \mission{ROSAT} and 2018 \mission{Swift/XRT} observations), the \eddratio is estimated to be less than a few $10^{-3}$, implying accretion via ADAF. While \eddratio increases by a factor of 100, to $\sim0.13$, at the high luminosity state, suggesting a thin accretion disc. Further evidence for a thin disc is supported by the significant detection of a soft component in the X-ray spectrum in the high luminosity state. The temperature ($\sim19^{+15}_{-7}\,\mathrm{eV}$) of this component is consistent with the predicted temperature of the inner region of a thin disc around a $\sim10^7$\msun. The time-scales of the accretion mode transitions are estimated to be several months to $\sim 1$ years from the multi-band data. Our results may indicate that the accretion mode transition can take place in AGN over a time-scale of months to a few years. Future intensive multi-band monitor of \ngc can possibly give a much better estimation of the time-scales at different transition stage which may help us understanding the physical process driven the accretion mode transition.


\section*{Acknowledgements}

ZL thanks Dr. Andy Beardmore for pointing out the effect of the bright earth on \mission{Swift/XRT} data, Professor Bifang Liu, Dr. Fuguo Xie, Dr. Chichuan Jin for helpful discussions and comments. This work is supported by the Strategic Pioneer Program on Space Science, Chinese Academy of Sciences, grant No. XDA15052100. WY and EQ acknowledge the support from the Strategic Priority Research Program of the Chinese Academy of Sciences grant No. XDB23040100. This work is also supported by the National Natural Science Foundation of China (grant no. 11673026, U1631238). EQ acknowledges supports from National Natural Science Foundation of China (grants no. 11773037). This work made use of data supplied by the UK Swift Science Data Centre at the University of Leicester. This work is based on observation obtained with \mission{XMM--Newton}, an ESA science mission with instruments and contributions directly fund by ESA Member States and NASA. This publication makes use of data products from the Wide-field Infrared Survey Explorer, which is a joint project of the University of California, Los Angeles, and the Jet Propulsion Laboratory/California Institute of Technology, and NEOWISE, which is a project of the Jet Propulsion Laboratory/California Institute of Technology. WISE and NEOWISE are funded by the National Aeronautics and Space Administration. Funding for the SDSS and SDSS-II has been provided by the Alfred P. Sloan Foundation, the Participating Institutions, the National Science Foundation, the U.S. Department of Energy, the National Aeronautics and Space Administration, the Japanese Monbukagakusho, the Max Planck Society, and the Higher Education Funding Council for England. The SDSS Web Site is \url{http://www.sdss.org}. The SDSS is managed by the Astrophysical Research Consortium for the Participating Institutions. The Participating Institutions are the American Museum of Natural History, Astrophysical Institute Potsdam, University of Basel, University of Cambridge, Case Western Reserve University, University of Chicago, Drexel University, Fermilab, the Institute for Advanced Study, the Japan Participation Group, Johns Hopkins University, the Joint Institute for Nuclear Astrophysics, the Kavli Institute for Particle Astrophysics and Cosmology, the Korean Scientist Group, the Chinese Academy of Sciences (LAMOST), Los Alamos National Laboratory, the Max-Planck-Institute for Astronomy (MPIA), the Max-Planck-Institute for Astrophysics (MPA), New Mexico State University, Ohio State University, University of Pittsburgh, University of Portsmouth, Princeton University, the United States Naval Observatory, and the University of Washington.



\bibliography{./references}
\bibliographystyle{mnras}






\bsp	
\label{lastpage}
\end{document}